\documentclass[journal]{IEEEtran}
\IEEEoverridecommandlockouts
\usepackage{amsmath}
\usepackage{amsfonts}
\usepackage{bbding}
\usepackage{amssymb}
\usepackage{array}
\usepackage{subfigure}

\usepackage{graphicx}
\usepackage{subfigure}
\usepackage[named]{algo}
\usepackage{algorithmic}
\usepackage{psfrag}
\usepackage{stfloats}
\usepackage[compress]{cite}
\makeatletter
\renewcommand{\citepunct}{,\penalty\@m\hskip.13emplus.1emminus.1em}
\renewcommand{\citedash}{\hbox{--}\penalty\@m}
\makeatother
\usepackage{setspace}
\usepackage{color}
\allowdisplaybreaks

\usepackage{amsthm}
\usepackage{stfloats}

\usepackage{hyperref}

\begin{document}
\title{Deep Learning for Ultra-Reliable and Low-Latency Communications in 6G Networks}

\author{
	\IEEEauthorblockN{{Changyang~She, Rui~Dong, Zhouyou~Gu, Zhanwei~Hou, Yonghui~Li, Wibowo Hardjawana,  Chenyang~Yang, Lingyang~Song, and Branka~Vucetic}}
	
\thanks{C. She, R. Dong, Z. Gu, Z. Hou, Y. Li, W. Hardjawana, and B. Vucetic are with the School of Electrical and Information Engineering, the University of Sydney, Sydney, NSW 2006, Australia (e-mail: \{changyang.she, rui.dong, zhouyou.gu, zhanwei.hou, yonghui.li, wibowo.hardjawana, branka.vucetic\}@sydney.edu.au). }
\thanks{C. Yang is with the School of Electronics and Information Engineering, Beihang University, Beijing 100191, China (email:cyyang@buaa.edu.cn).}
\thanks{L. Song is with the School of Electrical Engineering and Computer Science, Peking University, Beijing 100871, China (email:lingyang.song@pku.edu.cn).}
}

%

\maketitle
\begin{abstract}
In the future 6th generation networks, ultra-reliable and low-latency communications (URLLC) will lay the foundation for emerging mission-critical applications that have stringent requirements on end-to-end delay and reliability. Existing works on URLLC are mainly based on theoretical models and assumptions. The model-based solutions provide useful insights, but cannot be directly implemented in practice. In this article, we first summarize how to apply data-driven supervised deep learning and deep reinforcement learning in URLLC, and discuss some open problems of these methods. To address these open problems, we develop a multi-level architecture that enables device intelligence, edge intelligence, and cloud intelligence for URLLC. The basic idea is to merge theoretical models and real-world data in analyzing the latency and reliability and training deep neural networks (DNNs). Deep transfer learning is adopted in the architecture to fine-tune the pre-trained DNNs in non-stationary networks. Further considering that the computing capacity at each user and each mobile edge computing server is limited, federated learning is applied to improve the learning efficiency. Finally, we provide some experimental and simulation results and discuss some future directions.
\end{abstract}

\section{Introduction}
One of the most challenging objectives in beyond the 5th generation (5G) or so-called the 6th generation (6G) networks is to achieve ultra-reliable and low-latency communications (URLLC), which are the foundation for enabling many mission-critical applications with stringent requirements on end-to-end (E2E) delay and reliability \cite{Philipp2017Latency}. For example, autonomous vehicles, factory automation, and virtual/augmented reality (VR/AR) require a delay bound of $1\sim10$~ms and a packet loss probability of $10^{-5}\sim10^{-7}$ \cite{aijaz2019tactile}. Although the sum of the uplink and downlink transmission delays can be reduced to $1$~ms in the coming 5G New Radio, the randomness in wireless networks like dynamic traffic loads and uncertain channel conditions will result in serious network congestions. As a result, the E2E delay is much longer than the transmission delays in the air interface. Achieving the E2E delay and reliability requirements in such highly dynamic wireless networks presents unprecedented challenges in 6G networks.

The successful development of all the previous generations of mobile networks is mainly based on model-based methods, which are very useful in performance analysis and network optimization. However, the model-based methods alone cannot address the challenges in 6G networks. To obtain tractable results, some ideal assumptions and simplifications are inevitable in model-based methods. Thus, the obtained solutions cannot satisfy the quality-of-service (QoS) requirements in real-world networks. Moreover, the optimization problems for resource allocation, scheduler design, and network management are either non-deterministic or non-convex. To optimize the related policies according to dynamic parameters in wireless networks, the system needs to execute searching algorithms every few milliseconds. This will bring high computing overheads and long computing delay.

With recent advances in data-driven deep learning, it is possible to learn a wide range of policies for wireless networks \cite{sun2019application}. However, data-driven deep learning methods need a long training phase and a large number of training samples. To evaluate whether a policy satisfies the reliability requirement or not (e.g., packet loss probability of $10^{-5}\sim10^{-7}$), a device needs to send more than $10^5$ packets in the real-world network.
It takes a very long time to obtain enough training samples and will not be possible if the packet arrival rate of a device is low. To apply deep learning in URLLC, well-established models and theoretical formulas in communications and networking are helpful \cite{Dora2019Deep}. Merging model-based and data-driven methods is a promising approach in 6G networks.

In this paper, we first summarize some open issues when applying supervised deep learning or deep reinforcement learning in URLLC. Then, we develop a multi-level architecture that enables device intelligence, edge intelligence, and cloud intelligence for URLLC. In the architecture, model-based analysis results reveal the fundamental trade-off between latency and reliability in URLLC, and serve as the guidance and benchmarks for data-driven deep learning. Specifically, deep neural networks (DNNs) first learn from the optimal policies obtained from theoretical models and then transferred to real-world networks with deep transfer learning \cite{Chuanqi2018transfer}. To improve the learning efficiency at mobile users (MUs) and mobile edge computing (MEC) servers, federated learning is adopted in the architecture \cite{mcmahan2017communication}. Finally, we provide some experimental and simulation results and discuss future directions.



\section{Deep Learning in URLLC}
In this section, we summarize supervised deep learning and deep reinforcement learning (DRL) approaches that can solve non-deterministic problems and make decisions in real-time. Noting that applying deep learning in URLLC is not straightforward, we discuss some open problems of these methods.


\subsection{{DNN in Wireless Networks}}
{A policy in wireless networks can be described by a function that maps the network state to the decision on routing, scheduling, access control, resource allocation, and so on \cite{tang2019future}. Such a function is denoted by $y=f(x)$, where $x$ and $y$ are the state and the decision, respectively. In general, the closed-form expression of $f(\cdot)$ is hard to derive, and hence the system needs to search the optimal decision with when the state varies, e.g., channel states. Searching algorithms usually have high complexity and can not be implemented in real time.}

{To reduce the computing delay for executing searching algorithms, one can use a DNN, denoted by $\hat{y} = \Phi(x;\Theta)$, to approximate the optimal policy \cite{Haoran2018learning}, where $\Theta$ represents the parameters of the DNN and $\hat{y}$ is the output of the DNN with a given input $x$. By training the parameters, we can obtain an accurate approximation $\Phi(x;\Theta) \approx f(x)$. According to the universal approximation theorem, for a deterministic and continuous function $f(\cdot)$, the approximation error approaches to zero as the scale of the DNN increases \cite{Haoran2018learning}.}

\subsection{Supervised Deep Learning for URLLC}
{To apply supervised deep learning for URLLC, the system needs a large number of training samples, $(x_m,y_m), m=1,...,M$, which could be the optimal state-decision pairs obtained from an optimization algorithm or the historical data of traffic loads or trajectories of MUs.}

\subsubsection{Approximating optimal policies}  {By approximating optimal policies with DNNs, the computing delay for finding optimal solutions can be reduced remarkably. However, the approximation error will deteriorate the QoS of URLLC. To handle this issue, we should design a system conservatively. For example, when approximating the optimal resource allocation policy,} APs need to adjust the output of the DNN slightly by reserving a small portion of extra resources.

\subsubsection{Traffic and mobility predictions}
Traffic and mobility predictions have been exhaustively investigated in the existing literature. There are different kinds of prediction methods, such as time-series models, Markov chain models, and the Kalman filter. To apply the first two methods, we need some simplified traffic models or mobility models, which may not be accurate enough for URLLC. When applying the Kalman filter, one cannot exploit the long-term dependency of data. Compared with these prediction methods, recurrent neural networks can predict the traffic state from the historical data. For the data with long-term dependency, long-short-term-memory networks can be applied to further improve the performance.

%


\subsubsection{Open problems of supervised deep learning}
When applying supervised deep learning in URLLC, there are three major problems.
\begin{itemize}
\item \emph{Lack of labeled training samples.} To train a DNN, we need a large number of labeled training samples by solving optimization problems. Since optimization problems are non-convex in general. Obtaining a large number of optimal solutions is time-consuming.
\item \emph{Performance loss in non-stationary environments.} A DNN is usually trained offline. However, wireless networks are not stationary. As a result, the DNN can neither maximize the performance in terms of resource utilization efficiency nor guarantee the QoS of URLLC.
\item \emph{Prediction errors.} Predictions are not error-free. Prediction errors will lead to strong interference and high transmission collision probabilities \cite{Hou2018Burstiness}. However, for most of the deep learning algorithms, deriving the prediction error probability is very challenging.
\end{itemize}

\subsection{DRL for URLLC}\label{sub:DRL}
To address the above issues, one may turn to DRL since it does not need labeled training samples.

\subsubsection{Approximating the feedback of a decision} When applying reinforcement learning in URLLC, the dimension of the Q-table can be large, and it takes a very long time for the traditional Q-learning to converge, especially when the state space is continuous. To handle this issue, a DNN can be used to approximate the Q-function. For a given state and action, the Q-value can be obtained from the output of the DNN.

\subsubsection{Approximating optimal policies} If the policy in the DRL is deterministic, a DNN can also be used to approximate the optimal policy that maps the state of the system to the action. For example, the inter-slice resource management issue in network slicing was studied in \cite{qi2019deep}, where the resource management policy is approximated by a DNN.
\vspace{-0.0cm}
\begin{figure*}[tbp]
        \centering
        \begin{minipage}[t]{0.6\textwidth}
        \includegraphics[width=1\textwidth]{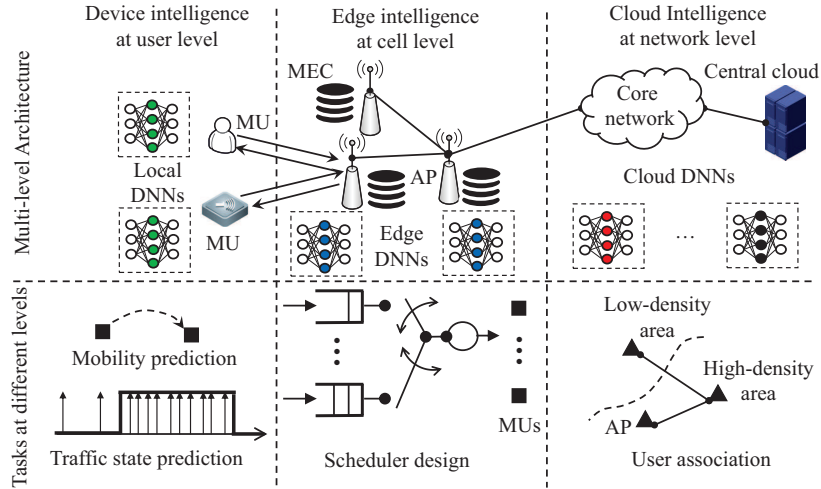}
        \end{minipage}
        \caption{Multi-level Architecture in 6G.}
        \label{fig:DT}
        \vspace{-0.6cm}
\end{figure*}

\subsubsection{Open problems of DRL} When applying DRL in URLLC, three issues remain unclear.

\begin{itemize}
\item \emph{Assumption on Markov decision process (MDP).} It is well-known that DRL can only be used to find optimal control policies of MDPs. However, problems in practical systems may not be Markovian.

\item \emph{Without QoS guarantee.} Unlike optimization problems that ensure QoS requirements by including some constraints, DRL does not have any constraint. Although we can design some heuristic rewards and punishments that take QoS into account, whether the achieved QoS can satisfy the requirements of URLLC is not clear \cite{qi2019deep}.
\item \emph{Exploration safety.} To improve the performance of the DRL, the system has to explore some unknown actions that may result in unexpected rewards or punishments. If the algorithm tries a bad action, the QoS requirement of URLLC cannot be satisfied.
\end{itemize}

\section{A Multi-level Architecture {in 6G}}
To address the open problems in the previous section, we propose a multi-level architecture that enables device intelligence, edge intelligence, and cloud intelligence at user level, cell level, and network level, respectively. In addition, deep transfer learning and federated learning are adopted in this architecture.

\subsection{{Features of 6G Networks}}
{The features of 6G Internet-of-Things networks and 6G vehicular networks are summarized in \cite{song2020artificial} and \cite{tang2019future}, respectively. To develop an architecture that enables deep learning for URLLC, the following features of 6G networks should be considered.}

\subsubsection{{E2E QoS requirement}} {When design 5G systems, we divided the system into multiple cascaded building blocks. As a result, the E2E latency and reliability requirements of URLLC can hardly be satisfied. In 6G networks, we need to ensure E2E QoS requirement by adjusting the whole network according to the stochastic service requests, queue states of buffers, workloads of servers, and wireless channels.}

\subsubsection{{Scalable and flexible control plane}} {With software-defined network, the control plane and user plane are slitted in 5G networks. For better scalability and flexibility in 6G networks, the network functions in the control plane could be fully centralized, partially centralized, or fully distributed \cite{song2020artificial}. Thus, the deep learning algorithms can be centralized or distributed depending on the network functions.}

\subsubsection{{Multi-level storage and computing resources}} {In 6G networks, storage and computing resources will be deployed at MUs, MEC, and central cloud \cite{tang2019future}. The central cloud has plenty of resources for offline training, but the communication delay between MUs and the cloud is long. With the help of MEC, it is possible to train DNNs locally, and the response time of the network is much shorter. By deploying computing resources at MUs, each device can make decision with it's local information in real time. Such a feature enables us to develop deep learning at different levels.}

\subsection{Multi-level Architecture}
{Based on the above features, we consider a wireless network that consists of smart MUs, MEC servers at APs, a central cloud in Fig. \ref{fig:DT}}. To better illustrate the multi-level architecture, mobility and traffic prediction for each MU, scheduler design at each AP, and user association in a multi-AP network are investigated.

\subsubsection{Device Intelligence at User Level} With device intelligence, MUs able to make decisions based on local predicted information, such as traffic state and mobility. Since the prediction reliability is crucial for making decisions, the prediction error probability should be extremely low. To analyze the prediction error probability, we can use a model-based method for prediction, and derive the prediction error probability \cite{hou2019prediction}. If data-driven methods outperform the model-based method, the prediction error probability achieved by the model-based method can serve as an upper bound of data-driven methods.

\subsubsection{Edge Intelligence at Cell Level}
A scheduler at an AP maps the channel state information and queue state information to resource allocation among different MUs. With edge intelligence, DRL can be used to optimize the scheduler. The basic idea is to use two DNNs to approximate the optimal scheduler and the value function, respectively.

With model-free DRL, the AP needs to evaluate the delay and reliability of a certain action by transmitting a large number of packets in the network. This leads to long convergence time of the DRL. To handle this issue, theoretical formulas can be used to evaluate the decoding error probability in the short blocklength regime and the queueing delay violation probability \cite{she2017cross}. Besides, by exploring in a numerical environment, the exploration safety can be improved remarkably \cite{qi2019deep}.

\subsubsection{Cloud Intelligence at Network Level}
User association schemes depend on the large-scale channel gains from MUs to APs as well as the packet arrival rate of each MU. With cloud intelligence, a centralized control plane uses a DNN to approximate the optimal user association scheme that maps the large-scale channel gains and the packet arrival rates of MUs to the user association scheme \cite{Dora2019Deep}. With a strong computing capacity, the central cloud can build a numerical platform that {mirrors} the behavior of the whole network. From the numerical platform, the system can explore labeled training samples with optimization algorithms, and then train the DNN with the optimal solutions. After the training phase, the DNN is saved at the control plane for online implementation.

Considering that real networks are highly dynamic, updating the states of all the network components to the central cloud will lead to unaffordable communication overheads. To avoid high overheads, the cloud only requires the information that is static or predictable in a large area or a long prediction horizon (hours). For example, the topology of APs, backhauls, and core networks are static and the density of MUs (or service requests) is predictable with spatial and temporal correlation. 

\subsection{Deep Transfer Learning in Non-stationary Environments}
{The theoretical formulas and models mentioned in the above architecture are used to initialize DNNs in stationary networks. However, real-world networks are non-stationary, and the models do not match the networks. Due to the model mismatch, the pre-trained DNN cannot guarantee the QoS.}


%
%

\begin{figure}[tbp]
        \centering
        \begin{minipage}[t]{0.5\textwidth}
        \includegraphics[width=1\textwidth]{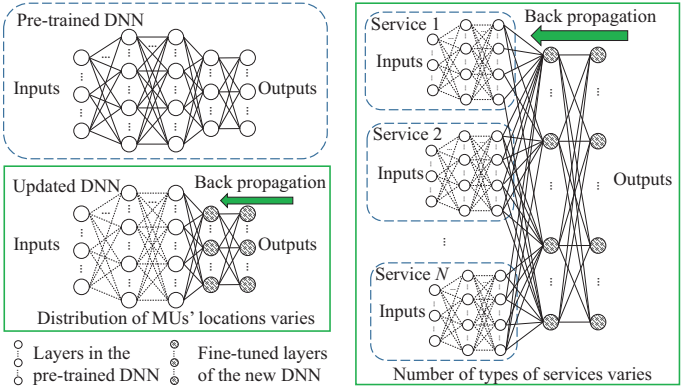}
        \end{minipage}
        \caption{Deep transfer learning.}
        \label{fig:Transfer}
        \vspace{-0.4cm}
\end{figure}

Deep transfer learning is a promising technique that can update the DNN with few training samples in non-stationary environments \cite{Chuanqi2018transfer}. The basic idea is reusing one part of the pre-trained DNN. As illustrated in the first example in Fig.~\ref{fig:Transfer}, when the distribution of MUs' locations varies, the system fine-tunes the last few layers of the DNN with new training samples. In the other example in Fig.~\ref{fig:Transfer}, the types of services change over time. A viable approach is to train a DNN for each type of services. When there are $N$ types of services in the network, the system changes {the structure of} the last few layers of the $N$ pre-trained DNNs and construct a larger DNN. Then, the system trains the last few layers of the DNN.

{It is worth noting that the output of the DNN cannot guarantee the QoS requirements in the re-training phase, during which the DNN is fine-tuned. To satisfy the QoS requirements, we still need optimization algorithms to find optimal solutions that are used as labeled training samples to fine-tune the DNN.}

\subsection{Federated Learning in the Multi-level Architecture}
To apply deep learning at MUs and MEC servers in the proposed architecture, there are two open problems. First, the number of local training samples may not be enough to train a DNN. Second, the computing capacity of an MU or an MEC is limited, thus if the DNN is trained locally, the training time will be long. A straight forward approach is to train DNNs at the central cloud by collecting data from all MUs and MEC servers. However, user data is privacy sensitive and large in quantity \cite{mcmahan2017communication}. This approach will lead to privacy issues and high communication overheads.

To enable device intelligence and edge intelligence, the edge-assisted hierarchical federated learning method proposed in \cite{liu2019edge} can be adopted in our multi-level architecture. As illustrated in Fig. \ref{fig:DT}, {MUs update parameters of local DNNs, $\Theta^{\rm L}_k, k=1,2,3, ...$, to MEC servers. The edge DNN at the $l$th MEC server, denoted by $\Theta^{\rm E}_l$, is obtained from the weighted sum of local DNNs, $\Theta^{\rm E}_l = \sum_k w^{\rm L}_k\Theta^{\rm L}_k$. The definitions of the weight coefficients can be found in \cite{liu2019edge}.} Then, the edge DNN is shared among all the MUs associated with this AP. Meanwhile, edge DNNs are sent to the central cloud periodically. {In the central cloud, the global DNN is obtained by aggregating the parameters of local DNNs, i.e., $\Theta^{\rm G} = \sum_l w^{\rm E}_l\Theta^{\rm E}_l$. Finally, the global DNN is shared to all MUs and APs.} With federated learning, the central cloud and MEC servers do not collect training samples from MUs, and hence will not cause privacy issue due to sharing data. Besides, the communication overheads for sharing DNNs are much lower than sharing data.

\section{Experiments and Simulation}
\begin{figure}[tbp]
        \centering
        \begin{minipage}[t]{0.35\textwidth}
        \includegraphics[width=1\textwidth]{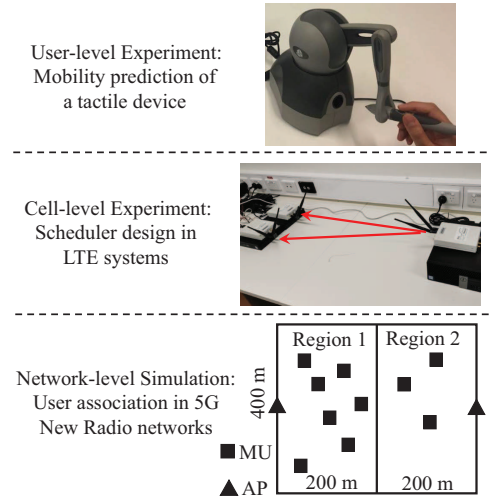}
        \end{minipage}
        \caption{Experiments and simulation at different levels.}
        \label{fig:DT2}
        \vspace{-0.4cm}
\end{figure}

In this section, we carry out some experiments and simulation to evaluate the performance of using deep learning in URLLC. As shown in the experiments and simulation in Fig.~\ref{fig:DT2}, we carry out user-level and cell-level experiments with a real tactile device and transceivers with Long Term Evolution (LTE) protocols. To further evaluate the performance in the coming 5G or 6G networks, we provide some network-level simulations with 5G New Radio.

\vspace{-0.0cm}

\subsection{User-Level Experiment: Mobility Prediction}
A real 3D System Touch tactile device is applied to control a virtual robotic arm. The location given by the device is updated every time slot with a duration of $1$~ms (i.e., the same as the transmission time interval in LTE systems), and is recorded for training and testing. The performance of the two prediction methods is evaluated. The first one is a model-based method based on Newton's laws of motion \cite{hou2019prediction}. The second method uses a fully-connected DNN to predict future locations from past locations. The reason why we use a fully-connected DNN is that it can achieve good performance for a small-scale prediction problem, and there is no need to use other types of DNNs. The inputs are the locations of the device in the past $50$~ms and the outputs are the predicted locations in the coming $20$~ms. There are two hidden layers, each of which has $100$ neurons. The parameters of the DNN are trained with $10^4$ training samples. After the training phase, the DNN is used to predict the mobility of the tactile device in an experiment with a duration of $2\times10^6$~ms.

\begin{table}[htbp]
\vspace{-0.2cm}\small
\renewcommand{\arraystretch}{1.3}
\caption{Prediction Error Probability}
\begin{center}\vspace{-0.4cm}\label{T:Prediction}
\begin{tabular}{|c|c|c|c|}
  \hline
  \multicolumn{4}{|c|}{Requirement on prediction accuracy $2$~cm} \\\hline
  Prediction horizon & $5$~ms & $10$~ms & $20$~ms  \\\hline
  Model-based& $6.61\times10^{-6}$ & $3.85\times10^{-5}$ & $3.20\times10^{-3}$ \\\hline
  Data-driven &  \multicolumn{3}{|c|}{$\ll 10^{-5}$}   \\\hline
  \multicolumn{4}{|c|}{Requirement on prediction accuracy $0.5$~cm} \\\hline
  Prediction horizon & $5$~ms & $10$~ms & $20$~ms  \\\hline
  Model-based &  \multicolumn{3}{|c|}{$\gg 10^{-5}$}   \\\hline
  Data-driven &  $4.59\times10^{-6}$ & $6.86\times10^{-6}$ & $2.25\times10^{-5}$  \\\hline
\end{tabular}
\end{center}
\vspace{-0.4cm}
\end{table}
The performance of model-based and data-driven methods for mobility prediction is shown in Table \ref{T:Prediction}, where the error probability is defined as the probability that the distance between the predicted location and the actual location is larger than the required prediction accuracy. The results in Table \ref{T:Prediction} show that it is possible to achieve high prediction reliability (i.e., $10^{-5}$ prediction error probability) with either model-based or data-driven methods, and the accuracy achieved by the data-driven method is better than that achieved by the model-based method. This is because the model used in the model-based method is not accurate enough to achieve high prediction accuracy \cite{hou2019prediction}.

\subsection{Cell-Level Experiment: Scheduler Design}
In the cell-level experiment, we apply an actor-critic DRL algorithm for scheduler design, where one AP serves two MUs. The radio transceivers are universal software radio peripheral B210. The AP has an Intel i7-8700 CPU with 6 cores and each MU is equipped with an Intel i7-6700 CPU with 4 cores. As discussed in Section \ref{sub:DRL}, two fully-connected DNNs are used to approximate the policy and the value function, respectively. The actor has two hidden layers, each of which has $40$ neurons. The critic also has two hidden layers, and each layer consists of $60$ neurons.

To reduce the training time and improve exploration safety, the actor and the critic are pre-trained in a simulation environment, where a digital model that mirrors the behavior of the real-world network is used to generate feedback to the DRL algorithm. The packet size is $200$~bytes, and the average packet arrival rate is $100$~packet/s. The total bandwidth is $5$~MHz. The DRL algorithm minimizes the overall packet loss probability subject to the requirements on delay and jitter, which are characterized by two delay bounds, $D_{\min}=9$~ms and $D_{\max}=11$~ms. The E2E delay should be higher than $D_{\min}$ and lower than $D_{\max}$, which means the jitter should be less than $2$~ms. To avoid long feedback delays and high jitters, retransmission is not allowed. 

The overall packet loss probabilities achieved by DRL in both the simulation and the real-world network are provided in Table~\ref{T:Scheduler}. To evaluate the packet loss probability in the simulation, both decoding errors and delay bound violations are taken into account. In addition to these two factors, hardware impairment in the real-world network will cause packet losses. The results in Table \ref{T:Scheduler} show that if the actor trained in the simulation environment is directly applied in the real-world system, the achieved reliability is worse than that evaluated in the simulation. After fine-tuning, the reliability can be improved, but is still worse than that in the simulation environment. There are two reasons: 1) the scheduler cannot control jitter caused by signal processing and data transmission in the practical system, while in the simulation environment this part of delay is fixed. 2) the modulation and coding scheme in LTE systems cannot achieve the minimum decoding error probability, which is computed from the Normal Approximation in simulation \cite{hou2019prediction}.

To reduce the user-experienced delay, we can combine user-level mobility prediction with cell-level scheduler design. According to the results in Table \ref{T:Prediction}, the prediction horizon can be up to $10$~ms with reliability better than $10^{-5}$. The latency evaluated in the real-world network lies in $[9,11]$~ms with packet loss probability around $10^{-2}$. If the AP sends the predicted locations to the MUs $10$~ms in advance, the user experienced delay will be $[-1,1]$~ms (a negative user experienced delay means that the user can predict the mobility of the transmitter). However, to improve the overall reliability, 5G New Radio and more advanced computing systems are needed.

\begin{table*}[htbp]
\vspace{-0.0cm}\small
\renewcommand{\arraystretch}{1.3}
\caption{Reliability achieved by DRL}
\begin{center}\vspace{-0.2cm}\label{T:Scheduler}
\begin{tabular}{|c|c|c|c|}
  \hline
   & Delay violation & Decoding error& Overall packet loss  \\\hline
  Evaluated in simulation environment & $2.05 \times 10^{-5}$ & $2.25 \times 10^{-4}$ & $2.46 \times 10^{-4}$ \\\hline
  Pre-trained actor in real-world system & $7.15 \times 10^{-2}$ & $5.73  \times 10^{-3}$ & $7.74 \times 10^{-2}$ \\\hline
  Fine-tuned actor in real-world system & $1.29 \times 10^{-2}$ & $4.55 \times 10^{-3}$ & $1.75 \times 10^{-2}$ \\\hline
\end{tabular}
\end{center}
\vspace{-0.2cm}
\end{table*}

\subsection{Network-Level Simulation: User Association}
We consider a wireless network with 5G New Radio, where two APs serve five delay-tolerant MUs and five URLLC MUs. The network topology can be found in the simulation and experiment of the network level in Fig. \ref{fig:DT2}. To reflect the impacts of non-stationary hidden variables on the performance of different schemes, we change the ratio of the number of MUs in Region~$1$ to that in Region~$2$ from $5:5$ to $9:1$ after $2000$ simulation trials.

Packets generated by each MU are either processed at the local server of the MU or offloaded to an MEC server. Since batteries of MUs have limited capacities, we minimize the maximal normalized energy consumption of MUs subject to QoS constraints. The normalized energy consumption of each MU is defined as the ratio of the energy consumption to the number of bits that have been processed. To train the DNN, $8000$ training samples explored from the numerical platform with the method in \cite{Dora2019Deep}. The DNN includes one input layer, one output layer, and four hidden layers, each of which consists of $100$ neurons. The inputs of the DNN include the large-scale channel gains and average packet arrival rates of all the MUs. The output of the DNN is the user association scheme. To illustrate the advantages of deep learning, it is compared with three baselines: 1) Each MU accesses to the AP with the highest large-scale channel gain (with legend `Highest SNR'). 2) A game theory approach developed in \cite{Jizhe2019Joint} (with legend `Game'), which is one of the most recent works on user association with hybrid services. 3) The optimal user association scheme obtained with the exhaustive searching method (with legend `Optimal').

\vspace{-0.0cm}
\begin{figure*}[t]
        \centering
        \begin{minipage}[t]{0.62\textwidth}
        \includegraphics[width=1\textwidth]{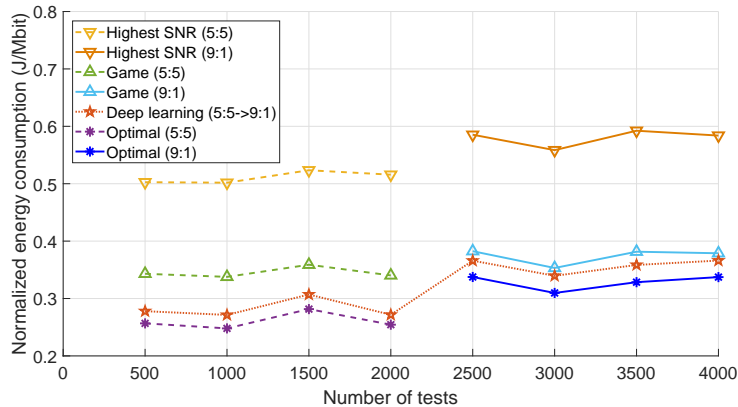}
        \end{minipage}
        \caption{Maximal normalized energy consumption with two APs.}
        \label{fig:results}
        \vspace{-0.2cm}
\end{figure*}

Simulation results in Fig. \ref{fig:results} show that the performance of the deep learning approach is better than the two existing schemes and is close to the optimal scheme. When the distribution of MUs locations changes, $500$ new training samples are used to fine-tune all the layers of the DNN. It is much smaller than the number of training samples that are needed to train a new DNN, e.g., $8000$ in our simulation.

\section{Future Directions}

\subsection{From Small-scale to Large-scale Networks}
As the numbers of devices and APs increase in the 6G networks, the number of parameters in the feed-forward DNN will be large. To avoid training a large number of parameters, one can use convolutional neural networks (CNNs). Since the number of parameters of a CNN does not increase with the dimension of the input, CNNs are very convenient in image processing. However, the topology of a wireless network is much more complicated than a two-dimensional image that can be represented by a matrix. To optimize radio resource allocation according to the topology of the network, we can use graph neural networks.

\subsection{From Centralized to Distributed Learning Algorithms}
With a global view of the wireless network, centralized learning algorithms can achieve better performance than distributed learning algorithms in terms of E2E delay, reliability, and resource utilization efficiency. However, centralized learning algorithms need to collect information from all the APs and MUs, and thus brings high overheads in backhauls and leads to long control-plane latency. With a distributed learning algorithm, each AP or MU can make its decisions according to local information. Nevertheless, how to guarantee the QoS requirement of URLLC with distributed algorithms deserve further study.

\subsection{From Wireless Networks to Interdisciplinary Research}
6G networks are expected to support applications in different vertical industries, such as vehicle networks, mission-critical Internet-of-Things, and VR/AR applications. The specific QoS requirements and traffic features of different applications are very different. Thus, interdisciplinary research is crucial for achieving the target requirements. By formulating theoretical models of both communication systems and specific applications in vertical industries, model-based methods enable us to understand, predict, and optimize the performance of the application from an interdisciplinary perspective. Based on the model-based analysis, we can design practical solutions with data-driven deep learning methods to approach fundamental limits and handle model mismatch problems.


\bibliographystyle{IEEEtran}
\bibliography{ref}

%
\begin{IEEEbiographynophoto}{Changyang She}
(S'12-M'17) received his B. Eng degree in Honors College (formerly School of Advanced Engineering) from Beihang University (formerly Beijing University of Aeronautics and Astronautics, BUAA), Beijing, China in 2012 and Ph.D. degree in School of Electronics and Information Engineering from BUAA in 2017. From 2017 to 2018, he was a postdoctoral research fellow at Singapore University of Technology and Design. Since 2018, he has become a postdoctoral research associate at the University of Sydney. His research interests lie in the areas of ultra-reliable and low-latency communications, tactile internet, mobile edge computing, internet-of-things, deep learning in 5G and beyond.
\end{IEEEbiographynophoto}

\begin{IEEEbiographynophoto}{Rui Dong}
received her M.S. degree with Distinction in School of Electrical \& Electronic Engineering, University of Manchester, the UK in 2012.  She is currently pursuing her Ph.D. degree in School of Engineering and Information Technologies, University of Sydney, Australia. Her research interests include beamforming on massive MIMO, user association design, resource allocation design and deep learning in wireless communications.
\end{IEEEbiographynophoto}

\begin{IEEEbiographynophoto}{Zhouyou Gu}
received his B.E. degree with First Class Honours and received his M.Phil. degree from the University of Sydney, Australia, in 2016 and in 2019. He is currently pursuing his Ph.D. degree in School of Electrical and Information Engineering at the University of Sydney, Australia. His research interests lie in the areas of programmable networks, wireless scheduler designs, deep reinforcement learning in 5G and beyond.
\end{IEEEbiographynophoto}

\begin{IEEEbiographynophoto}{Zhanwei Hou}
received B.E. degree in electrical engineering from Sun Yat-sen University (formerly Zhongshan University) in 2011, M.E. degree in computer science from institute of computing technology, Chinese academy of sciences in 2014, and Ph.D. degree from University of Sydney in 2019. He is currently a postdoctoral research associate in the University of Sydney. His research interests are Tactile Internet, ultra-reliable and low-latency communications, and industrial Internet of Things.
\end{IEEEbiographynophoto}

\begin{IEEEbiographynophoto}{Yonghui Li}
(M'04-SM'09-F'19) received his PhD degree in November 2002 from Beijing University of Aeronautics and Astronautics. Since 2003, he has been with the Centre of Excellence in Telecommunications, the University of Sydney, Australia. He is now a Professor in School of Electrical and Information Engineering, University of Sydney. He is the recipient of the Australian Queen Elizabeth II Fellowship in 2008 and the Australian Future Fellowship in 2012.

His research interests are in the area of wireless communications, with a particular focus on MIMO, millimeter wave communications, machine to machine communications, coding techniques and cooperative communications. He holds a number of patents granted and pending in these fields. He received the best paper awards from IEEE International Conference on Communications (ICC) 2014, IEEE PIMRC 2017 and IEEE Wireless Days Conferences (WD) 2014. He is Fellow of IEEE.
\end{IEEEbiographynophoto}

\begin{IEEEbiographynophoto}{Wibowo Hardjawana}
(M'09) received the Ph.D. degree in electrical engineering from The University of Sydney, Australia, in 2009. He was an Australian Research Council Discovery Early Career Research Award Fellow and is now Senior Lecturer with the School of Electrical and Information Engineering, The University of Sydney. Prior to that he was with Singapore Telecom Ltd. His current research interests are in cellular radio access and wireless local area networks, with focuses in system architectures, resource scheduling, signal processing and the development of corresponding standard-compliant prototypes.
\end{IEEEbiographynophoto}

\begin{IEEEbiographynophoto}{Chenyang Yang}
(M'99-SM'08) received her Ph.D. degrees from Beihang University (BUAA), China, in 1997. She has been a full professor with the School of Electronics and Information Engineering, BUAA since 1999. She has published over 200 papers in the fields of energy-efficient transmission, URLLC, wireless local caching, CoMP, interference management, cognitive radio, and relay, etc. She was supported by the 1st Teaching and Research Award Program for Outstanding Young Teachers of Higher Education Institutions by Ministry of Education of China. She was the chair of Beijing chapter of IEEE Communications Society during 2008-2012,. She has served as TPC Member, TPC co-chair or Track co-chair for many IEEE conferences. She has ever served as editors for IEEE Trans. on Wireless Communication, IEEE Journal of Selected Topics in Signal Processing and IEEE Journal of Selected Areas in Communications. Her recent research interests lie in mobile AI, wireless caching, and URLLC.
\end{IEEEbiographynophoto}

\begin{IEEEbiographynophoto}{Lingyang Song}
(S'03, M'06, SM'12, F'19) received his Ph.D. degree from the University of York, United Kingdom, in 2007, where he received the K. M. Stott Prize for excellent research. He worked as a research fellow at the University of Oslo, Norway, until joining Philips Research UK in March 2008. In May 2009, he joined the School of Electronics Engineering and Computer Science, Peking University, and is now a Boya Distinguished Professor. His main research interests include wireless communication and networks, signal processing, and machine learning. He was the recipient of the IEEE Leonard G. Abraham Prize in 2016 and the IEEE Asia Pacific (AP) Young Researcher Award in 2012. He has been an IEEE Distinguished Lecturer since 2015.
\end{IEEEbiographynophoto}

\begin{IEEEbiographynophoto}{Branka Vucetic}
(SM'00-F'03) is currently an ARC Laureate Fellow and the Director of the Centre of Excellence for IoT and Telecommunications, The University of Sydney. Her current research work is in wireless networks and Internet of Things. In the area of wireless networks, she focuses on the communication system design for millimeter-wave frequency bands. In the area of Internet of Things, she focuses on providing wireless connectivity for mission-critical applications.

Ms. Vucetic is a fellow of the Australian Academy of Technological Sciences and Engineering and the Australian Academy of Science.
\end{IEEEbiographynophoto}

%

\end{document}